\documentclass[a4paper]{jpconf}

\usepackage{amsmath,amsfonts,amssymb}
\usepackage{graphicx}
\usepackage[colorlinks=true, allcolors=blue]{hyperref}

\usepackage{color}
\usepackage[latin9]{inputenc}
\usepackage{mathrsfs,amsmath}
\usepackage{graphicx}%
\usepackage{float}
\usepackage{amsfonts}%
\usepackage[titletoc]{appendix}
\usepackage{amssymb}
\usepackage{braket}
\usepackage{bm}
\usepackage{xr}

\usepackage[T1]{fontenc}

\def\p{\partial}

\newcommand{\includegraphicsXL}[1]{\includegraphics[width = 0.80\textwidth]{#1}}

\newcommand{\reffig}[1]{Fig.~\ref{#1}}
\newcommand{\refeq}[1]{Eq.~(\ref{#1})}

\newcommand{\refsec}[1]{Sec.~\ref{#1}}

\begin{document}
\title{Passive and hybrid mode locking in multi-section terahertz quantum cascade lasers}

\author{P. Tzenov}
\address{Technical University of Munich, 80333 Munich, Germany}
\author{I. Babushkin}
\address{Institute of Quantum Optics, Leibniz University Hannover, 30167 Hannover, Germany}
\address{Max Born Institute, 12489 Berlin, Germany}
\author{R. Arkhipov}
\address{St. Petersburg State University,199034 St. Petersburg, Russia} 
\address{ITMO University,  197101 St. Petersburg, Russia} 
\author{M. Arkhipov}
\address{St. Petersburg State University,199034 St. Petersburg, Russia} 
\author{N. Rosanov}
\address{ITMO University,  197101 St. Petersburg, Russia} 
\address{Vavilov State Optical Institute, Kadetskaya Liniya v.o. 14/2, St Petersburg 199053, Russia}
\address{Ioffe Physical Technical Institute, Politekhnicheskaya str. 26, St Petersburg 194021, Russia}
\author{U. Morgner}
\address{Institute of Quantum Optics, Leibniz University Hannover, 30167 Hannover, Germany}
\author{C. Jirauschek}
\address{Technical University of Munich, 80333 Munich, Germany}

\ead{petar.tzenov@tum.de}

\begin{abstract}
	It is believed that passive mode locking is virtually impossible in
	quantum cascade lasers (QCLs) because of too fast carrier relaxation
	time. Here, we revisit this possibility and theoretically show that 
	stable mode locking and pulse durations in the few cycle regime at terahertz
	(THz) frequencies are possible in suitably engineered bound-to-continuum QCLs.
	We achieve this by utilizing a multi-section cavity geometry with alternating gain and absorber sections.
	The critical ingredients are the very strong coupling of the absorber to both field and environment as well as a fast absorber carrier recovery dynamics. Under these conditions, even if the gain relaxation time is several times faster than the cavity round trip time, generation of few-cycle pulses is feasible. We investigate three different approaches for ultrashort pulse generation via THz quantum cascade lasers, namely passive, hybrid and colliding pulse mode locking. 
	\end{abstract}

\section{Introduction}
\label{sec:introduction}

Quantum cascade lasers (QCLs) are unipolar, electrically pumped semiconductor devices in which the optical transition occurs between bound 
electron states in the conduction band of a specially designed quantum well heterostructure~\cite{williams2007terahertz}. Due to the 
intersubband nature of the radiative transition, QCLs are highly tunable and allow for the generation of coherent radiation in the 
underdeveloped terahertz (THz) and mid-infrared (MIR) portions of the electromagnetic spectrum. 

Since the first experimental realization of a QCL in 1994 \cite{faist1994quantum}, this technology has experienced remarkable advancement,
with some of the most notable milestones being the realization of a room temperature MIR QCL emitting power at the Watt level \cite{bai2008room}, 
the demonstration of a THz QCL operating at the record high temperature of $~$200 Kelvin \cite{fathololoumi2012terahertz}, as well as the 
successful generation of broadband coherent frequency combs by free running devices both in the MIR and THz spectral regions
\cite{hugi2012mid,burghoff2014terahertz}.   

Naturally, it is also of great scientific and practical interest to enable the formation of short, mode locked pulses of light with QCLs. This
would be a major advancement for THz and MIR spectroscopy as it will open up the stage for ultrafast optical experiments, such as for example
time-resolved THz spectroscopy \cite{ulbricht2011carrier}, with compact, on-chip, direct sources. Additionally, since mode locked pulses are
frequency combs in the Fourier domain, ultrashort pulse generation via QCLs will provide an alternative approach to obtain broadband frequency
combs. 

Unfortunately, experience shows that QCLs are notoriously difficult to mode lock \cite{wang2015generating}, with the shortest pulse widths
achieved so far being around 2.5 ps in the THz via active modulation of the injection current~\cite{bachmann2016short}. It is believed that, due
to the ultrafast processes that govern intersubband transitions, active mode locking of QCLs is feasible only close to lasing threshold, whereas
passive mode locking (PML), in the traditional sense, is virtually impossible \cite{wang2009mode}. This is because the intrinsically short carrier 
relaxation times, typically several times smaller than the cavity round trip time, obstruct the formation of short bursts of light since the 
trailing edges of any propagating pulse would be amplified by the fast recovering gain \cite{haus2000mode}. 

We believe that there is no fully conclusive evidence to support these claims, especially in the THz, as the gain recovery dynamics has not been
extensively studied. In fact, to the our best knowledge, to present date there have been only two publications experimentally investigating the 
gain recovery time in bound-to-continuum (BTC) THz devices, and none in resonant-phonon QCLs. Interestingly, both experimental results indicate 
sub-threshold lifetimes on the order of several tens of picoseconds \cite{green2009gain,bacon2016gain}. These measurement techniques are based 
on a pump-probe experimental method where a perturbing resonant pump pulse is injected into the gain medium followed by a temporally detuned 
probe pulse interacting with the saturated gain. In the publication in Ref.~\cite{green2009gain}, the photocurrent induced by stimulated emission
between the upper and the lower laser level was recorded as a function of the delay between both pulses, and a Gaussian fit was used to infer the
speed of the recovery of population in the upper laser state. The measured lifetimes were  $\approx 50$ ps which, as we will show, are long
enough to enable mode locking. In fact one might argue that BTC QCLs are amongst the most optimal devices for mode locking, since the energy
exchange between the propagating pulse and the saturable gain is most efficient when the carrier dynamics is faster than the round trip time. 

Our idea is based on well established techniques for quantum dot and conventional semiconductor lasers \cite{avrutin2000monolithic,vladimirov2005model, rafailov2005high, rafailov2007mode,arkhipov2013hybrid,arkhipov2016self},
where PML is routinely achieved based on a saturable absorber (SA) and a gain medium as separate components of a multi-section wave guide. Here absorption is implemented by reverse biasing the gain medium. 
Carrying this concept to QCLs was first suggested by Franz K{\"a}rtner \cite{kaertner08}, whereas Talukder and Menyuk  were
the first to point out that rather than reverse biasing the gain, carefully chosen positive biases should be used for QCLs~\cite{talukder2014quantum}. In this work the authors simulated passive mode locking for MIR devices, 
where the intensity dependent saturation was implemented via a (slow) quantum coherent absorber. Along these lines simulations for PML in THz QCLs have also been 
presented~\cite{tzenov2017passive,talukder2017ultra}. Here, we expand upon the previous work by i) using an extended theoretical model, ii) showing that guided by classical principles one
 can achieve passive mode locking by simply using a fast saturable absorber instead~\cite{haus1975theory}, and iii) in addition to the conventional PML approach, 
 we also discuss the possibility of hybrid and colliding pulse mode locking.

The way in which the fast saturable absorber (FSA) enables mode locking is two-fold~\cite{jones1995dynamics,williams2004long}. First, the FSA provides more gain for shorter pulses,
 strong enough to bleach the material, while at the same time it also suppresses weak background fluctuations. Secondly, it also acts as a 
compensator for the dispersion introduced by the gain medium, as both gain and loss interact resonantly with the intracavity intensity, albeit
with different signs in the polarization term. As a result, if the gain and absorber sections are packed into a compact structure, with the
small round trip time only several times longer than the relaxation time in the gain section, very stable mode locking with one or two pulses
per round trip arises. 

This paper is organized as follows: in \refsec{sec:model} we present the theoretical model and in \refsec{sec:modelocking} we investigate several different 
approaches which might lead to the generation of picosecond THz pulses via QCLs, namely conventional passive mode locking, \refsec{subsec:PML}, colliding pulse 
mode locking, \refsec{subsec:CPML}, and hybrid mode locking, \refsec{subsec:HML}. 

\section{Theoretical model}
\label{sec:model}

\begin{figure}[h!]
	\centering
	\includegraphicsXL{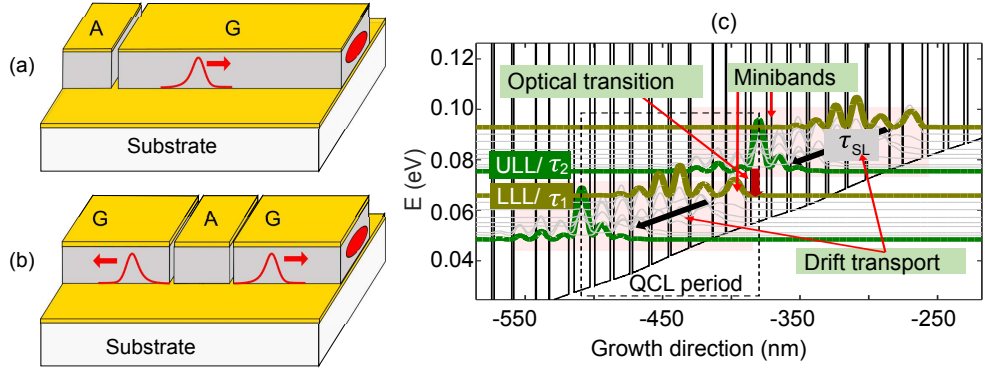}
	\caption{An example of multi-section Fabry-Perot (FP) cavity geometry, consisting of spatially separated gain (G) and absorber (A) 
		regions for (a) conventional passive mode locking and (b) colliding pulse mode locking. (c) Conduction band diagram and wave functions of a typical bound to continuum (BTC) GaAs/Al$_{0.15}$GaAs$_{0.85}$ THz quantum
		cascade laser (the structure is analogous to that in Ref.~\cite{barbieri20042}). The upper and lower laser levels, ULL and LLL, are 
		outlined with thick lines. The electron transport in the device is characterized by drift transport (scattering) inside the miniband 
		and optical transition between minibands.}
	\label{fig:cavity}
\end{figure}

The multi-section cavity design envisaged by us is illustrated in  \reffig{fig:cavity} for a Fabry-Perot (FP) geometry and two different
configurations, the A-G and G-A-G alignments, favouring conventional and colliding pulse mode locking, respectively. The general geometry 
consists of two or more sections with different biases and effective dipole moments. This can be realized either via wafer-bonding of separately 
designed and grown structures, or by designing a single heterostructure operating either as a gain or absorber medium,
 depending on the driving current \cite{schwarz2014monolithically}. 
Despite the more challenging fabrication as compared to having epitaxially stacked gain and absorption layers, or external cavity multi-section QCLs~\cite{revin2016active},  we insist on monolithic wave guides as they offer two obvious advantages: i) these
structures provide short round trip lengths, i.e. relatively small round trip times, and also ii) arranging the gain and absorber in series,
as depicted in the figure, allows for independent control of the injection current in all sections. 

Before we  write down the equations of motion, a careful consideration of the transport processes in a BTC quantum cascade laser
is in order. An exemplary such active region is illustrated in Fig.~\ref{fig:cavity}(c). Typically, electron transport through the heterostructure
can be described by three different lifetimes, the superlattice relaxation time, i.e. $\tau_{SL}$, defined as the transit time of a carrier from
the top of the miniband to the upper laser level, $\tau_{2}$ defined as the lifetime of the upper laser level and lastly $\tau_{1}$ denoting the 
same for the lower laser level~\cite{choi2008gain}. 

A usual modelling approach in literature is to eliminate the population of the lower laser level, $\rho_{11}$, from the system of
equations by assuming that $\rho_{11} \approx 0$ at all times~\cite{talukder2014quantum,gordon2008multimode,jirauschek2014modeling}. This choice can be justified by the relatively fast out-scattering from this
level to lower energetic states in the miniband, compared to the other non-radiative lifetimes in the system, 
i.e.  $\tau_1 \ll \tau_2,\tau_{SL}$. Such an approximation is valid for determining the steady state solutions of the rate equations in the 
absence of an optical field, however it breaks down when one additionally considers photon assisted scattering, since it dramatically reduces the 
upper laser level lifetime~\cite{choi2008gain}. Concretely, when $1/\tau_1 < 1/\tau_2+1/\tau_{st}(|E|^2)$, where $\tau_{st}(|E|^2)$ is the stimulated
emission/absorption lifetime and $E$ is the electric field, the dynamics of the lower laser level can no longer be excluded, since the QCL 
essentially  operates as a three level system~\cite{choi2008gain}. A value of $\tau_1\approx 2$ ps was reported for the first THz BTC-QCL~\cite{kohler2002terahertz},
 indicating that typical values for the lower laser level lifetime are of that order. 

Keeping this in mind, we employ a density matrix model to describe the electron transport through the triplet
$\rho_{SL}$, $\rho_{22}$ and $\rho_{11}$ for the population density of the electrons in the miniband, the upper laser level and the lower laser level,
respectively. Where necessary, we denote the various system parameters with sub-/superscript index $g$ to indicate that those quantities 
are related to the gain section alone. A similar system was used in the work of Choi et al. to provide evidence for quantum coherent 
dynamics in MIR QCLs, where an excellent agreement between simulation and experimental data was achieved~\cite{choi2010ultrafast}.

Expanding on the usual two level Bloch equations approach, we write down the Maxwell-Bloch (MB) equations for  
the three level system, in the rotating wave and slowly varying amplitude approximations, taking into account counter-propagating waves and 
spatial hole burning. The full system of equations for the gain medium is given by	
\begin{subequations}
	\label{eq:dynamics_G}
	\begin{align}
	& \frac{\p E_{\pm}^g}{\p x} \pm \frac{n_0}{c} \frac{\p E_{\pm}^g}{\p t} = -i\frac{\Gamma_g \mu_g\omega_0}
	{\varepsilon_0 c n_0}N_g\eta_{\pm}-\frac{a}{2}E_{\pm}^g, \label{eq:dynamics_G:field}\\
	& \frac{d\rho_{SL}^{0}}{dt} = \frac{\rho_{11}^{0}}{\tau_{1}} - \frac{\rho_{SL}^{0}}{\tau_{SL}}  
	\text{  ,} \\
	& \frac{d\rho_{22}^{0}}{dt} = \frac{\rho_{SL}^{0}}{\tau_{SL}} - \frac{\rho_{22}^{0}}{\tau_{2}} +
	i\frac{\mu_{g} }{2\hbar} \big [ (E_{+}^g)^* \eta_{+} + (E_{-}^g)^* \eta_{-}   - c.c. \big], \\
	& \frac{d\rho_{11}^{0}}{dt} = \frac{\rho_{22}^{0}}{\tau_{2}} - \frac{\rho_{11}^{0}}{\tau_{1}}  - 
	i\frac{\mu_{g}  }{2\hbar} \big [ (E_{+}^g)^* \eta_{+} + (E_{-}^g)^* \eta_{-}   - c.c. \big], \\
	& \frac{d\rho_{SL}^{+}}{dt} = \frac{\rho_{11}^{+}}{\tau_{1}} - \frac{\rho_{SL}^{+}}{\tau_{SL}}  \text{  ,}  \\
	& \frac{d\rho_{22}^{+}}{dt} = \frac{\rho_{SL}^{+}}{\tau_{SL}} - \frac{\rho_{22}^{+}}{\tau_{2}} 
	+ i\frac{ \mu_{g}  }{2\hbar} \big [ (E_{-}^g)^* \eta_{+} -
	(E_{+}^g)\eta_{-}^* \big], \\
	& \frac{d\rho_{11}^{+}}{dt} = \frac{\rho_{22}^{+}}{\tau_{2}} -\frac{\rho_{11}^{+} }{\tau_{1}} - i\frac{ \mu_{g}  }{2\hbar} \big [ (E_{-}^g)^* \eta_{+} - 
	(E_{+}^g)\eta_{-}^* \big], \\	
	& \frac{d \eta_{\pm}}{dt}= -i(\omega_{g}-\omega_0)\eta_{\pm}+i\frac{ \mu_{g} }{ 2 \hbar} \big [ E_{\pm}^g (\rho_{22}^{0}
	-\rho_{11}^{0}) + E_{\mp}^g(\rho_{22}^{\pm}-\rho_{11}^{\pm}) \big ] - \frac{\eta_{\pm}}{T_{2g}} \label{eq:dynamics_G:eta}  \text{.}
	\end{align}
\end{subequations}
The symbols $E_{\pm}^{g}$ denote the forward and backward propagating field envelopes and $\eta_{\pm}$ the corresponding slowly varying
coherence terms between levels 1 and 2. The interference pattern of the counter-propagating waves leads to the formation of standing waves inside the cavity, which 
results in a population grating and consequently in spatial hole burning. According to the standard approach~\cite{gordon2008multimode,wang2007coherent,gkortsas2010dynamics},
we take the following ansatz for the population of the $j^{\text{th}}$ state 
\begin{equation}
\rho_{jj} = \rho_{jj}^{0} + \rho_{jj}^{+}e^{2ik_0 x} + (\rho_{jj}^{+})^* e^{-2ik_0x},
\end{equation}
where  $\rho_{jj}^{0}$ is the average population, $\rho_{jj}^+$ denotes the (complex) amplitude of the grating, $k_0 = \omega_0 n_0/c$ is the
carrier wave number expressed in terms of the carrier angular frequency $\omega_0$ and the background refractive index is $n_0\approx 3.6$. Furthermore, $c$ denotes the velocity of light in vacuum, $\varepsilon_0$ the 
permittivity of free space and $\hbar$ the Plank constant. The rest of the simulation parameters for both absorber and gain are specified in Table~\ref{tab:params}.

In order to impose a minimal set of assumptions about the absorber, we model it as a two level density matrix system in the rotating wave and slowly varying amplitude approximation with 
average inversion $\Delta^0$, inversion grating amplitude $\Delta^+$ and a coherence term $\pi_{\pm}$.
Again, we use sub-/superscript $a$ to specify where a particular parameter or variable relates solely to the saturable absorber. The usual MB equations read
\begin{subequations}
	\label{eq:dynamics_A}
	\begin{align}
	& \frac{\p E_{\pm}^a}{\p x} \pm \frac{n_0}{c} \frac{\p E_{\pm}^a}{\p t} = -i\frac{\Gamma_a\mu_a\omega_0}{\varepsilon_0 c n_0}N_a 
	\pi_{\pm}-\frac{a}{2}E_{\pm}^a,\label{eq:dynamics_A:field}\\
	& \frac{d\Delta^{0}}{dt} =  i\frac{\mu_a }{\hbar} \big [ (E_{+}^a)^* \pi_{+} + (E_{-}^a)^* \pi_{-}   - c.c. \big] - 
	\frac{\Delta^0-\Delta^{\text{eq}}}{T_{1a}},\label{eq:dynamics_A:inversion} \\
	& \frac{d\Delta^{+}}{dt} =  i\frac{ \mu_a }{2\hbar} \big [ (E_{-}^a)^* \pi_{+} - (E_{+}^a)\pi_{-}^* \big] - 
	\frac{\Delta^{+}}{T_{1a}}  \text{  ,} \\
	& \frac{d \pi_{\pm}}{dt}= -i(\omega_{a}-\omega_0)\pi_{\pm}+i\frac{ \mu_a }{ 2 \hbar} \big [ E_{\pm}^a \Delta^{0} 
	+ E_{\mp}^a\Delta^{\pm} \big ] - \frac{\pi_{\pm}}{T_{2a}}\label{eq:dynamics_A:pi} \text{.}
	\end{align}
\end{subequations}

By far the most well established method for the numerical analysis of mode locking in semiconductor lasers is the travelling wave model~\cite{williams2004long,yang1993study, jones1995dynamics}, which treats the optical field in a similar manner as in \refeq{eq:dynamics_G:field} and \refeq{eq:dynamics_A:field}, however restricts the modelling of the gain/absorber dynamics to classical rate equations. This essentially "flat gain" approximation necessitates the inclusion of additional numerical techniques to 
impose the bandwidth limit of the gain medium~\cite{williams2004long}. By contrast, the Maxwell-Bloch equations intrinsically capture the spectral dependence of the gain, namely via 
the inclusion of the polarization equations, i.e. \refeq{eq:dynamics_G:eta} and \refeq{eq:dynamics_A:pi}, and thus constitute a more complete model. 

\begin{table}[h!]
	\centering 
	\caption{
		The parameters for the absorber (A) and gain (G) section of the
		two-section ring QCL from \reffig{fig:cavity}(a). In the tables below $e\approx 1.602\times 10^{-19}$C
		denotes the elementary charge.}
	\label{tab:params}
	\begin{tabular}{c|c|c|c}
		\hline
		Parameter & Unit & Value (G) & Value (A) \\
		\hline
		Dipole matrix el. ($\mu_{j}$) & nm$\cdot$e &  2  & 6 \\ 
		Resonant angular freq. ($\omega_{j}$)  & ps$^{-1}$ & $3.4\times2\pi$ & $3.4\times2\pi$ \\
		Gain superlattice transport time ($\tau_{SL}$) & ps & 40 & $\times$ \\ 
		Gain upper laser level lifetime  ($\tau_{2}$) & ps & 40 & $\times$ \\
		Gain lower laser level lifetime  ($\tau_{1}$) & ps & 2 & $\times$\\
		Absorber  lifetime ($T_{1a}$) & ps & $\times$  & 3 \\  
		Dephasing time ($T_{2a/g}$) & fs & 200 & 160 \\  
		Length ($L_j$)  & mm & 1 & 0.125 \\
		Doping density ($N_j$)  & cm$^{-3}$  &5$\times 10^{15}$ & 1$\times10^{15}$ \\
		Overlap factor ($\Gamma_j$) &dimensionless& 1.0 & 1.0  \\ 
		Linear power loss ($a$)  & cm$^{-1}$ & 10 & 10 \\
		\hline
	\end{tabular}
\end{table}

\section{Modelocking of QCLs}
\label{sec:modelocking}

In the following sections we investigate various schemes which might enable the generation of ultrashort pulses with THz QCLs. Besides conventional passive mode locking,
we also treat colliding pulse and hybrid mode locking as alternative approaches to improve the pulse characteristics. Importantly, for our envisaged design to work,  a 
slowly saturable gain, with inversion recovery time only several times faster than 
the round trip time, must be coupled with a fast saturable absorber~\cite{williams2004long}. To model this scenario we assume a parameter set as presented in Table \ref{tab:params}, with values we believe realistic for THz QCLs. Specifically, for the BTC-QCL lifetimes assumed in Table~\ref{tab:params}, simulations similar to Ref.~\cite{talukder2011modeling} yield a gain inversion lifetime of $T_{1g}\approx 12$ ps, while for the 1.125 mm FP cavity the round trip time is about 28 ps.  In all of the following sections, we present results from simulations of free-running, self-starting devices. To solve \refeq{eq:dynamics_G} and \refeq{eq:dynamics_A}, we use the numerical method outlined in~\cite{tzenov2016time} and start all simulations from random noise. Due to the nature of the employed approximations, our results are limited to pulses with durations not significantly shorter than $\sim 1$ ps. 
For sub-picosecond dynamics,  memory effects become also relevant, and can be taken into account by using a non-Markovian approach, however at the cost of considerably increased numerical complexity~\cite{butscher2005ultrafast, knezevic2013time}. 

\subsection{Passive mode locking (PML)} \label{subsec:PML}

One of the main results from the classical theory of passive mode locking is the condition that the absorber should saturate faster than the gain~\cite{haus1975theory}.
The non-linear saturation parameter is given by $\epsilon_j= \mu_j^2 T_{1j}T_{2j}/\hbar^2$ and denotes the inverse of the saturation 
value of the electric field squared $|E|^2$ in each active region ($j=\{a,g\}$). When the condition $r = \epsilon_a /\epsilon_g > 1 $ is met, the 
propagating pulse will bleach the absorber more strongly than the amplifier and thus will open a net round trip gain window. In fact, simulations for quantum dot lasers have shown \cite{williams2004long} that
the pulse duration decreases approximately exponentially with increasing value of $r$. Conversely, classical theory and also our simulations (results not shown here) predict that
 no mode locking is possible when $r<1$ \cite{williams2004long,haus1975theory}.
\begin{figure}[h!]
	\centering
	\includegraphicsXL{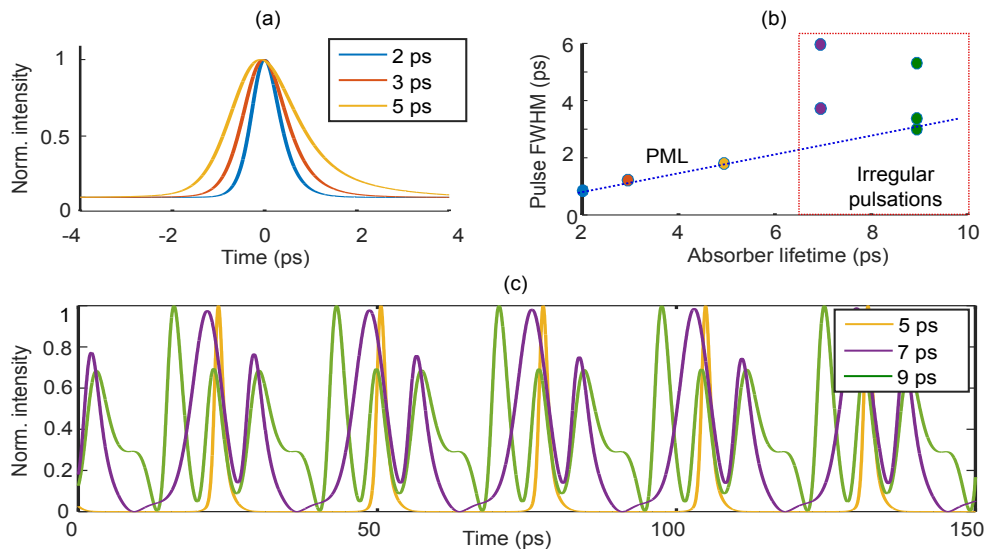}
	\caption{(a) Normalized pulse intensity vs time for values of the absorber lifetimes of 2 ps, 3 ps and 5 ps. (b) The intensity FWHM duration as  $T_{1a}$ is varied between 2-9 ps, where for $T_{1a}= 7$ ps and 9 ps, the FWHM values of the main and also the satellite pulses are presented. (c) The normalized optical intensity for $T_{1a}=$5-9 ps.}\label{fig:PML}
\end{figure}

Another requirement for successful mode locking is that the absorber should have a fast recovering population inversion. One can easily see
the benefits of short $T_{1a}$ lifetime for mode locking. Upon  entering the absorber, the pulse front will saturate the active medium, which
on the other hand, will be quick enough to recover prior to arrival of the pulse tail. This type of dynamics would naturally shorten the
pulse as the duration of the net gain window will decrease with decreasing $T_{1a}$. Following this logic, one might expect to obtain shorter
pulses with decreasing absorber lifetimes, which is indeed confirmed by our simulations. Importantly, absorbers with fast carrier recovery ought to be easy to realize based on resonant phonon QCL designs, taking advantage of strong longitudinal optical phonon scattering.

Similarly to their zero-dimensional counterparts (quantum dot lasers), we argue that QCLs could be passively mode locked provided systems with suitably chosen 
parameters are designed. To illustrate this possibility we simulated \refeq{eq:dynamics_G} and \refeq{eq:dynamics_A} with a parameter set 
characteristic for QCLs (see Table~\ref{tab:params}). 

To investigate the importance of absorber lifetime for PML of QCLs, we varied the recovery time of the absorber between 2 and 9 ps and simulated
the coupled system for around 400 round trips. Since the change of $T_{1a}$ also changes the value of $r$, for each simulation
we re-adjusted the absorber dipole moment in order to maintain constant $r$. This was necessary since we wanted to have controlled numerical experiments
where only $T_{1a}$ and not $r$ was varied. Finally, the gain carrier density was also adjusted from its value in Table~\ref{tab:params}, in order to ensure that in all subsequent simulations
 the active medium was biased at 1.2 times above threshold. 

The results from these simulations are presented in \reffig{fig:PML}(a) and \reffig{fig:PML}(b) and display behaviour in agreement with our expectation.
When the absorber lifetime is sufficiently short, \reffig{fig:PML}(a), as $T_{1a}$ increases so does also the pulse duration. In fact, for the PML regime of lasing
in \reffig{fig:PML}(b) we observe a clear linear relationship between $T_{1a}$ and the intensity full width at half maximum (FWHM) pulse duration. On the other hand, for slower absorbers, the 
complicated interplay between the optical field and the active region dynamics produces irregular pulsations (IP) with no well defined temporal profile. From \reffig{fig:PML}(c) we see that the onset of this regime occurs already for absorber lifetimes $T_{1a}\geq 7$ ps and is characterized by multiple pulses with varying intensity. Additionally, for those cases one can also observe modulation of the pulse amplitude with a period spanning several tens of round trips, a phenomenon baring resemblance to Q-switched mode locking~\cite{brovelli1995control}.  These results unequivocally validate the important role of the absorber lifetime for the pulsation dynamics and confirm that for successful mode locking of QCLs, besides slowly saturable gain media, also absorbers with short $T_{1a}$ lifetime and large $r=\epsilon_a/\epsilon_g$ ratio are essential.

\subsection{Colliding pulse mode locking (CPML)} \label{subsec:CPML}
\begin{figure}[h!]
	\centering
	\includegraphicsXL{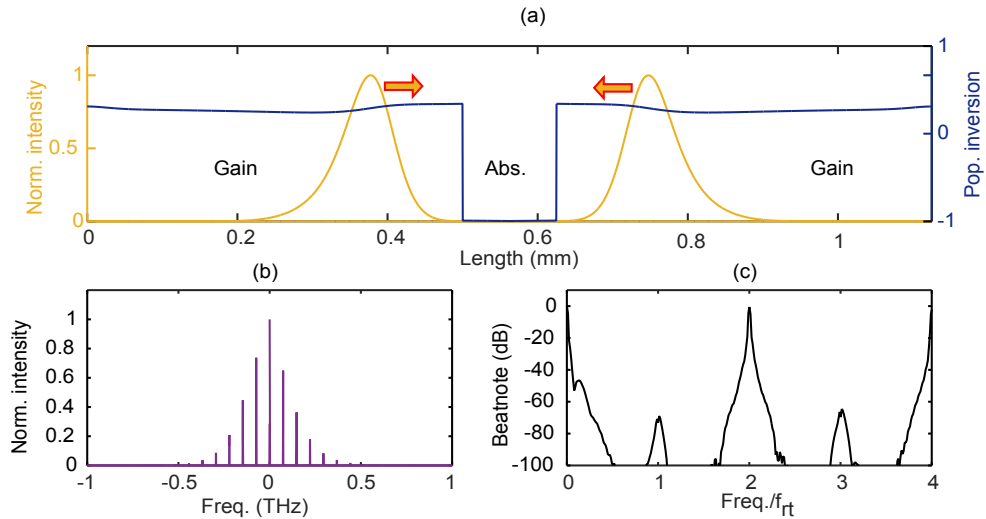}
	\caption{(a) A snapshot of the optical intensity (left y-axis) and population inversion (right y-axis) inside the cavity. (b) Optical spectrum of the field emitted from the 
		right facet of the cavity. (c) A log plot of the beatnote signal versus frequency (normalized to the cavity round trip frequency $f_{rt}$) produced by the device.}
	\label{fig:CPML}
\end{figure}

A special type of passive mode locking, useful for shortening even further the pulse duration, but probably more importantly to achieve high repetition rates, 
is the so called colliding pulse mode locking, where two gain sections of equal length are symmetrically placed around the absorber \cite{chen1992monolithic,jones1995dynamics}. When such a geometric arrangement is achieved, two identical pulses per round trip can be emitted 
from the device, resulting in doubled repetition rate equal to the second harmonic of the round trip frequency. In fact, we expect that
CPML will be easier to achieve via BTC quantum cascade lasers as the short gain recovery time will naturally favour such 
multi-pulse regime of operation \cite{tzenov2017passive}. 

To understand why CPML occurs, consider the schematic in \reffig{fig:cavity}(b), illustrating a multi-section cavity design in the G-A-G (gain-absorber-gain) configuration.
Let us assume that a single pulse with amplitude $E_0$ propagates inside a gain medium with some group velocity $v_g$. Close to the resonator
mirrors, during its forward pass the pulse will saturate the gain and there will be not enough time for the latter to recover in order to
re-amplify the reflected signal. This leads to a  reduction of the effective length of the gain medium by $\Delta L = v_g \tau_{gr}/2$, which is
half the distance travelled by the pulse in  time $\tau_{gr}$, where $\tau_{gr}\neq T_{1g}$ denotes the time it takes for the gain to recover 
to its threshold value. It can be shown that $\tau_{gr}$ is a monotonously increasing function of $|E_0|^2$, and as such $\Delta L$ will be shorter if the 
pulse would split into two identical copies with \emph{half} the total intensity each, since then $\tau_{gr}$ will also decrease. The most stable two-pulse configuration in a G-A-G Fabry-Perot cavity are indeed pulses,
colliding in the cavity centre, as those will saturate the absorber more deeply and further reduce the round trip losses.

To confirm these expectations, we simulated \refeq{eq:dynamics_G} and \refeq{eq:dynamics_A} in the G-A-G arrangement for the parameter
set in Table \ref{tab:params}. Figure~\ref{fig:CPML} illustrates the results. After about 50 round trips 
 the laser emission transforms into two identical counter-propagating pulses which collide inside the centre of the cavity, \reffig{fig:CPML}(a). The spectrum in \reffig{fig:CPML}(b) consists of more than 15 modes 
separated by twice the round trip frequency, $f_{rt}$, whereas beatnote calculations, \reffig{fig:CPML}(c), indicate a strong component at the second harmonic of $f_{rt}$. This second harmonic regime is stable over hundreds of round trips, with the beatnote linewidth in \reffig{fig:CPML}(c) being limited by the Fourier transform resolution of our simulations. Such a device essentially represents a very stable local oscillator with a repetition frequency of around $\approx$71 GHz~\cite{sonnabend2005evaluation}.

\subsection{Hybrid mode locking (HML)} \label{subsec:HML}
\begin{figure}[h!]
	\centering
	\includegraphicsXL{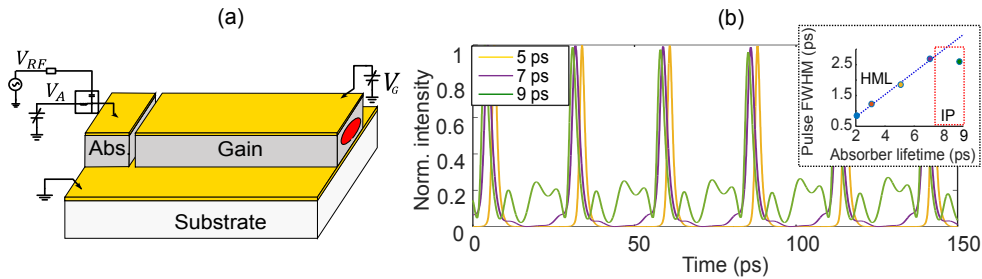}
	\caption{(a) Schematic diagram of hybrid mode locking with RF modulation of the absorber bias. (b) Intensity profile for $T_{1a}=$2-5 ps. (Inset) Intensity FWHM vs absorber lifetime for the hybrid configuration of (a) and the same parameter set as the one which produced the results in \reffig{fig:PML}(b). }
	\label{fig:HML}
\end{figure}
\noindent
As shown above, the multi-section gain/saturable absorber system might lase in a regime of irregular pulsations~\cite{jones1995dynamics}. In order to enforce phase-locking of the modes, one can additionally modulate the current/voltage
of the gain or the absorber at the round trip frequency of the optical field. Such a hybrid approach could be envisaged for QCLs,
where the applied bias of the absorber is superimposed with a sinusoidally varying radio frequency (RF) source via a bias-T, and the gain medium is pumped with a DC current
(see \reffig{fig:HML}(a))~\cite{arkhipov2013hybrid,fiol2010hybrid}. 

To simulate this technique we introduced the term $m_0 \sin (2\pi f_{rt} t)$ in the right hand side of \refeq{eq:dynamics_A:inversion},
modelling the RF source~\cite{gkortsas2010dynamics}, where the modulation amplitude was set to 25 \% of the DC current, i.e. $m_0 = 0.25 \times \Delta_{a}^{eq}/T_{1a}$.
Again, we repeated the simulations from \reffig{fig:PML} to evaluate how effective  this active+passive mode locking will be as compared to the simple PML case. The results are plotted in \reffig{fig:HML}(b). 

From \reffig{fig:HML}(b), within the fast absorber regime (i.e. $T_{1a} = 2-5$ ps),
we see that applied RF modulation does not seem to have any significant impact on the pulse widths, as the calculated FWHM-values are of almost the same magnitude as in \reffig{fig:PML}(b). This is not so surprising as the laser already operates in a
mode locking regime and so additional RF-injection would have little to no effect on the dynamics. However, substantial improvement in the pulse structure and duration is observed for slower absorbers as the satellite pulsations are strongly suppressed in favour of more regular pulses. In particular, comparing the time-domain profile of the intensity in \reffig{fig:PML}(c) and \reffig{fig:HML}(b), we see that as a result of this additional active modulation, the pulse at $T_{1a}=7$ ps has completely recovered its integrity whereas the pulse substructures for $T_{1a}=9$ ps are drastically reduced. Again, drawing insights from the quantum dot laser community~\cite{derickson1991comparison,thompson200310}, one might expect that with this HML technique an improvement in the overall stability of the pulse train and mode locking parameter range can be achieved, as compared to when utilizing the passive mechanism alone.

\section{Conclusion} \label{sec:conclusion}
We have suggested feasible approaches for ultrashort pulse generation in self-starting bound-to-continuum terahertz quantum cascade lasers. Our scheme is based on the realization of a
paradigmatic model for passive mode locking via a fast saturable absorber, implementable via multi-section monolithic Fabry-Perot cavities. We predict the formation of short picosecond pulses with FWHM limited by the gain bandwidth of the device. Our investigations show that besides a suitably engineered gain medium with slowly recovering population inversion, a fast saturable absorber with very strong coupling to the optical field is essential. Carefully conducted simulation experiments indicate that the multi-section configuration is prone to entering into a regime of irregular pulsations if the absorber recovery time is very large, which should be an important point to consider in future designs. Furthermore, besides passive mode locking, we have also discussed alternative approaches to ultrashort pulse generation in QCLs, i.e. hybrid and colliding pulse mode locking. By utilizing active modulation of the injection current in the absorber, the former method recovers the regular pulsations from a regime of irregular such. On the other hand, CPML might be easier to achieve with THz QCLs as multi-pulse lasing is the naturally preferred mode of operation in fast gain recovery active media.  

\section*{Funding Information}
This work was supported by the German Research Foundation (DFG) within the Heisenberg program (JI 115/4-1) and under DFG Grant No. JI 115/9-1 and the Technical University of Munich (TUM) in the framework of the Open Access Publishing Program.

\end{document}